\documentclass[12pt]{iopart}
\usepackage{iopams}
\usepackage{amssymb}

\usepackage{epsf,graphicx,rotating,color}
\usepackage{bm,bbm}

\newcommand{\la}{\langle}
\newcommand{\ra}{\rangle}

\newcommand{\eg}{{\it e.g.}}

\begin{document}

\title[Simulation of Grover's quantum search algorithm]{Simulation of Grover's 
   quantum search algorithm in a Ising nuclear spin chain quantum computer with 
   first and second nearest neighbour couplings}
\author{G V L\'opez, T Gorin and L Lara}
\address{Departamento de F\'{\i}sica, Universidad de Guadalajara\\
   Blvd. Marcelino Garc\'{\i}a Barragan y Calzada Ol\'{\i}mpica\\
   4480 Guadalajara, Jalisco, M\'exico}
\eads{gorin@pks.mpg.de}

\begin{abstract}
We implement Grover's quantum search algorithm on a nuclear spin chain quantum 
computer, taking into Ising type interactions between nearest and second 
nearest neighbours into account. The performance of the realisation of the 
algorithm is studied by numerical simulations with four spins. We determine
the temporal behaviour of the fidelity during the algorithm, and we compute
the final fidelity as a function of the Rabi frequency. For the latter, we 
obtained pronounced maxima at frequencies which fulfil the condition of the
$2\pi k$-method with respect to the second nearest neighbour interactions.
\end{abstract}

%PACS:03.67.-a,   % Quantum information
%     03.67.Lx,   % Quantum computation
%    03.65.Ta,   % Foundations of quantum mechanics; measurement theory
%    03.67.Dd,   % Quantum cryptography
%    03.67.Hk    % Quantum communication
%    03.67.Pp    % Quantum error correction and other methods for protection
                 % against decoherence

%\submitto{\JOB}

\maketitle

\section{\label{I} Introduction}

The Ising-spin chain quantum computer is a simple theoretical model system,
which has many of the relevant features of a physical realisation of a 
quantum computer. The bare model is defined in terms of a time-independent 
Hamiltonian $H_0$, describing a system of spins in a strong magnetic field and 
coupled by the Ising-interaction. Additional time-dependent perturbations
(\eg\ radio-frequency pulses, which may be switched on and off at will) are
used to implement certain elementary quantum gates, from which any unitary
quantum algorithm can be constructed (universal quantum gates).

Originally, this model has been proposed 
in~\cite{Lloyd00,Ber94,Lloyd01,Bennet00}, and a large number of theoretical 
studies has been devoted to it (to mention a few important 
works:~\cite{Ber00,Ber01b,BerDoo02}). As a physical realisation, one
may think of nuclear spins embedded in a solid state system in an external
strongly inhomogeneous magnetic field. Ultracold atoms in optical lattices 
may provide a possibly less demanding alternative~\cite{GRCi03}.
 
In the present work, we assume an Ising-interaction between nearest and 
next-nearest (second) neighbours~\cite{LoLa1,LoGoLa1}. In this case, the bare 
Hamiltonian becomes
\begin{equation}
\frac{1}{\hbar}\; H_0= -\sum_{k=1}^n w_k\; I^z_k 
 - 2J\sum_{k=1}^{n-1} I^z_k\, I^z_{k+1} - 2J'\sum_{k=1}^{n-2} I^z_k\, I^z_{k+2}
 \; , 
\label{I:defH0}\end{equation}
where $w_k$ is the Larmor frequency of spin $k$. Since the dipole interactions
may be induced by electron clouds around the nuclear spins, it is realistic to
consider $J$ and $J'$ as independent parameters. The Hamiltonian $H_0$ is 
diagonal in the so called computational basis, which consists of products
of single spin (qubit) eigenstates of $I^z_k$, which are denoted by $|0_k\ra$
and $|1_k\ra$ respectively:
\begin{equation}
I^z_k\; |\alpha_k\ra = \frac{(-1)^{\alpha_k}}{2}\; |\alpha_k\ra \; .
\end{equation}
We denote the $2^n$ eigenstates of $H_0$ by 
$\{|\alpha_{n-1},\cdots,\alpha_0\ra\}$ with 
$\alpha_k=0,1$ for $k=0,\cdots,n-1$.

In order to implement the desired quantum gates, we use monochromatic 
electromagnetic radio-frequency pulses (RF-pulses) described by the term 
$W(t)$,
\begin{equation}
H= H_0 + W(t) \; ,\qquad
W(t)=  \frac{\hbar\Omega}{2}\sum_{k=o}^{n-1}\left( \rme^{\rmi (wt+\varphi)}
   I_k^+ + \rme^{-\rmi (wt+\varphi)} I_k^-\right) \; ,
\label{I:defH}\end{equation}
where the frequency $w$, the phase offset $\varphi$ and the Rabi frequency
$\Omega$ are free parameters ($\Omega/\gamma$ is the amplitude of the 
electromagnetic field and $\gamma$ is the gyromagnetic ratio for the spins).
$I_k^+ = (I_k^-)^\dagger= |0_k\ra\la 1_k|$ is the raising operator of qubit
$k$. The desired quantum gates (and the whole quantum algorithm) are realised 
within the interaction picture. We apply the above defined RF-pulses with a 
rectangular envelope, choosing the frequency $w$ in resonance with an 
appropriate transition in $H_0$. The resulting unitary evolution is denoted by
\begin{equation}
R_k^{\mu\nu}(\varphi,\vartheta) \; ,\qquad w= w_k + \mu J + \nu J'\; ,
\label{I:Rres}\end{equation}
where $\mu$ and $\nu$ are integers. The angle $\varphi$ gives the phase offset 
of the RF-pulse, whereas $\vartheta= \Omega\tau$ determines the duration $\tau$
of the pulse. The RF-pulses are selected in such a way that they yield the
desired quantum gates exactly, if all off-resonant transitions are neglected
(resonant approximation). For the ``real'' evolution will be obtained from
numerical simulations, where all transitions are taken into account. The 
deviation from the ``ideal'' evolution is quantified by fidelity~\cite{fidrev}, 
the absolute value squared of the wave-function overlap for both evolutions. In
order to minimise the decay of fidelity, We apply the $2\pi k$-method to 
suppress those transitions which are nearest to resonance. These are of order
$J'$, and therefore related to the second neighbour interaction~\cite{LoGoLa1}.

In the present work, we study the effects of unitary errors (due to 
non-resonant transitions) on a quantum algorithm of intermediate length.
Such an algorithm may consist of several hundreds of pulses, even though it is 
still far away from the regime, where a quantum computer could really 
demonstrate its superiority over a classical one (after all, our work relies
on numerical simulations on a classical PC). It is however also far beyond 
pulse sequences, which only realise one or a few elementary qubit gates. Of 
particular interest will be the question whether we can from the fidelity 
decay of few gate operations extrapolate to the fidelity decay of the whole 
quantum algorithm. 

As an example we consider Grover's search algorithm ~\cite{Gro1,Gro2} which 
has also been implemented experimentally~\cite{Yang1,Feng1,Jon1}. We will 
implement this algorithm on a spin chain of length four, where one of the spins
will serve as the ``ancilla'' qubit. The basic elements of this algorithm are 
single-qubit Hadamard gates. Due to the Ising interactions, these gates are 
quite expensive in the Ising-spin chain quantum computer as we will see below. 
The big majority of the pulses are used to implement these gates. The other 
gates are $n$-qubit phase-gates, where $n$ is the number of bits of the search 
space (in our case $n=3$).

In the following section (section~\ref{G}), we will discuss Grover's search 
algorithm as such, and its implementation in the Ising-spin chain quantum 
computer. In section~\ref{S}, we show numerical simulations of the search 
algorithm, discuss the fidelity decay during the execution of the 
algorithm, and study this fidelity as a function of Rabi frequency. 
Conclusions are provided in section~\ref{C}.

\section{\label{G} Realisation of Grover's search algorithm}

In its most basic form~\cite{NieChu00}, Grover's quantum search algorithm 
requires $n$ qubits to prepare the ``enquiry'' states (in the data register)
subsequently presented to the ``oracle'', and a single additional qubit (the 
``ancilla'' qubit) required by the oracle to communicate its
answer. Starting from the ground state of the system, $|0_{n-1},\cdots,0_0\ra$,
the algorithm itself consists of an initial step, where the complete
superposition of all basis states is created by applying Hadamard gates to all
qubits of the data register. Let us denote a single Hadamard gate applied to
qubit $j$ by $H_j$ and the sequence of Hadamard gates to create the 
superposition state in $m$ qubits by $H^{\otimes m}$. The creation of the 
superposition state is followed by a Grover operator, which may be repeated 
several times, until the data register contains the desired target state with 
sufficiently high probability.

The Grover operator is again composed of Hadamard gates and conditional phase
reflections:
\begin{equation}
G= O(\alpha)\; H^{\otimes m}\; S_0\; H^{\otimes m} \; ,
\end{equation}
where $\alpha$ may be any number representable in the data register. The
unitary oracle $O(\alpha)$, leaves all basis states untouched, except for the
state $|\alpha\ra$, to which it applies a sign change (phase reflection). The
other unitary operator $S_0$, in turn, changes the sign of all basis states but
leaves the state $|0\ra$ untouched. While the Hadamard gates are single qubit 
gates, $O(\alpha)$ and $S_0$ are true $n$-qubit gates. 

In this work we implement the Grover algorithm in a quantum register of four 
qubits: $\{i_3,i_2,i_1,i_0\ra\}$. For numerical simulations, it is convenient 
to use the qubits $k= 0,2,3$ as data register, and the qubit $k=1$ as the 
ancilla qubit. In this way, the ancilla qubit is coupled directly (by first or
second neighbour couplings) to all qubits in the data register. In what follows, 
we discuss the implementation of the Hadamard gates, the conditional phase 
reflection $S_0$ and the oracle, in separate subsections.

\subsection{Hadamard gates}

The Hadamard gate is a single qubit gate. In the Hilbert space of that qubit
the gate acts as a unitary matrix:
\begin{equation}
H= \frac{1}{\sqrt{2}}\left(\begin{array}{cc} 1 & 1\\ 1 & -1\end{array}\right) 
\; .
\end{equation}
It can be decomposed into elementary qubit rotations as follows:
\begin{equation}
H= R_y(\pi)\; R_y(\pi/2)\; R_x(\pi) \; ,
\label{G:Hgen}\end{equation}
where $R_x(\vartheta)= R(\pi,\vartheta)$, $R_y(\vartheta)= R(\pi/2,\vartheta)$
and
\begin{equation}
R(\varphi,\vartheta)= \left(\begin{array}{cc} 
   \cos(\vartheta/2) & \rmi\,\rme^{\rmi\varphi}\; \sin(\vartheta/2)\\
   \rmi\,\rme^{-\rmi\varphi}\; \sin(\vartheta/2) & \cos(\vartheta/2)
\end{array}\right) \; .
\label{G:Rgen}\end{equation}
The single qubit rotations of the form~(\ref{G:Rgen}) may be obtained by 
applying appropriate electromagnetic pulses as defined in~(\ref{I:defH}) 
and~(\ref{I:Rres}) to the Ising spin chain. In the presence of neighbouring 
qubits, one has to repeat that pulse sequence for all possible configurations 
of the neighbouring qubits, in order to obtain the true single qubit gate,
independent of the state of the neighbouring qubits. Denoting the Hadamard gate 
applied to qubit $i$ as $H_i$, we obtain specifically:
\begin{eqnarray}
H_0 = \prod_{\mu= -1,1} \prod_{\nu= -1,1} R_0^{\mu\nu}(\pi/2,\pi)\;
   R_0^{\mu\nu}(\pi/2,\pi/2)\; R_0^{\mu\nu}(\pi,\pi) \label{G:Hada0}\\
H_2 = \prod_{\mu= -2,0,2} \prod_{\nu= -1,1} R_2^{\mu\nu}(\pi/2,\pi)\;
   R_2^{\mu\nu}(\pi/2,\pi/2)\; R_2^{\mu\nu}(\pi,\pi) \\
H_3 = \prod_{\mu= -1,1} \prod_{\nu= -1,1} R_3^{\mu\nu}(\pi/2,\pi)\;
   R_3^{\mu\nu}(\pi/2,\pi/2)\; R_3^{\mu\nu}(\pi,\pi) \; ,
\end{eqnarray}
which implements the unitary gate $H^{\otimes 3}= H_0 H_2 H_3$ on the
data register.

\subsection{The conditional phase inversion $S_0$}

Formally, the gate $S_0$ may be described by
\begin{equation}
S_0\; |\alpha\ra = (-1 + 2\delta_{\alpha,0})\; |\alpha\ra \; .
\end{equation}
The (unconditional) phase inversion of the state of a single qubit may be 
obtained by applying the pulse $R(0,2\pi)$ to it. In the present
case, we obtain $S_0$ by applying $R(0,2\pi)$ to the qubit $i=1$ repeatedly
-- covering all possible configurations of its neighbouring qubits except for
that case, where the data-qubits form the state $|0\ra$. This leads to the
pulse sequence
\begin{equation}
S_0=  R_1^{2,-1}(0,2\pi)\; R_1^{0,1}(0,2\pi)\; R_1^{0,-1}(0,2\pi)\;
   R_1^{-2,-1}(0,2\pi)\; R_1^{-2,1}(0,2\pi) \; .
\end{equation}

\subsection{The oracle}

The oracle is a unitary gate, which changes the sign of the state of the data
register, depending on its contents. In the search
algorithm, it is thought that any number $|\alpha\ra$ representable in the data
register serves as an index to given items in a data base. The search consists
in finding the index of that item which has a certain unique property (\eg\
the telephone number of a certain person with a certain address). In the 
present case, we will request that the index itself be a certain number
out of the set $\{0,5,8,13\}$ (decimal notation). In those cases, the oracle 
may be implemented by a single $2\pi$-pulses:
\begin{eqnarray}
O(0) = R_1^{2,1}(0,2\pi)\\
O(8) = R_1^{2,-1}(0,2\pi) \\
O(5) = R_1^{-2,1}(0,2\pi)\\
O(13) = R_1^{-2,-1}(0,2\pi)\; .
\end{eqnarray}
The other numbers representable in the data register are $\{1,4,9,12\}$, but
a oracle selecting those states would require many more qubits.

\section{\label{S} Numerical simulations}

Our simulations normally start with the ground state,
$|0000\ra$. Then, we generate the total superposition state in the data
register and perform two Grover steps. This is sufficient to obtain the desired
target state with very high probability (above 95\%). With the pulse
sequences given in section~\ref{G}, one needs 42 pulses for the initial 
superposition state and 90 pulses for each Grover operator step.

Unfortunately, the pulse sequence described in section~\ref{G} is not optimal in 
terms of accuracy. The direct implementation of the Hadamard gates according 
to the scheme described there, leads to prohibitively large errors. Luckily,
these errors can be reduced efficiently, by recombining the pulses in the
sequence for one Hadamard gate into shorter sequences implementing several
single qubit gates. This can be done due to the commutativity of the operators 
appearing in (\ref{G:Hada0}). Thus, for \eg\ the  Hadamard gate $H_0$, we 
obtain the following pulse sequence
\begin{equation}
\fl H_0= \left\{ \textstyle 
   \prod_{\mu,\nu=-1,1} R_0^{\mu\nu}(\pi/2,\pi) \right\}
   \left\{ \textstyle    
   \prod_{\mu,\nu=-1,1} R_0^{\mu\nu}(\pi/2,\pi/2) \right\}
   \left\{ \textstyle    
   \prod_{\mu,\nu=-1,1} R_0^{\mu\nu}(\pi,\pi) \right\} \; ,
\end{equation}
with much smaller errors. We use similar rearrangements for $H_2$ and $H_3$.

The overall error can be further decreased by rearranging the outer sequences
of $\pi$-pulses in a similar manner into sequences of single qubit 
$\pi/2$-pulses. These measures drastically increase the number of pulses for
the implementation of the algorithm. Now, we need 20 pulses for the $H_0$ 
and the $H_3$ gate and 30 pulses for the $H_2$ gate. In total this gives 70 
pulses for the $H^{\otimes 3}$ gate and 146 pulses for each Grover operator 
step. Note however that the amount of time to implement those gates is the 
same as before.
%  70           70
% 146          164
% 146          164
%----          ---
% 362 pulses   398 pi/2 steps

\begin{figure}
\input{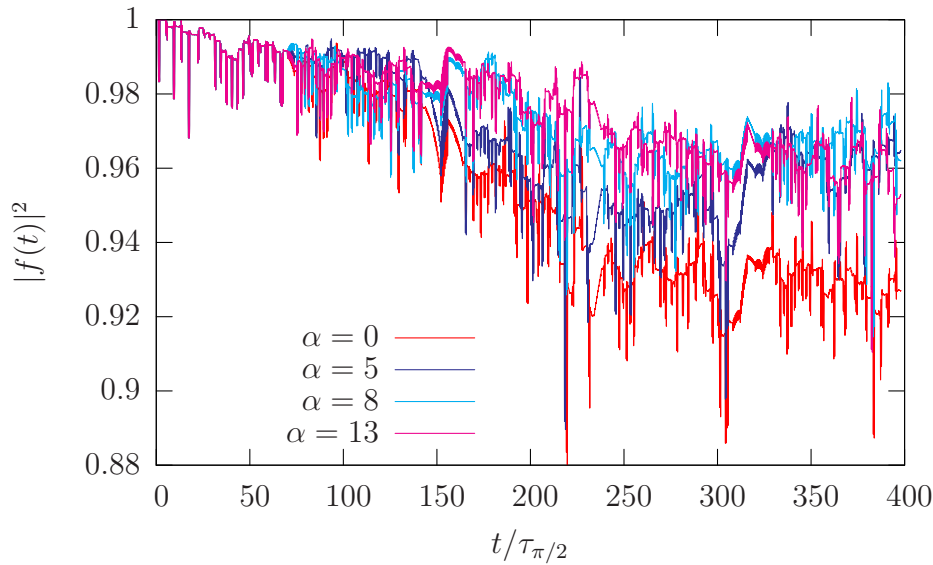}
\caption{The absolute value squared of the fidelity during the Grover 
algorithm with two Grover operator steps. Time is measured in units of the 
duration of a single $\pi/2$-pulse. The algorithm is simulated for different 
target states $|\alpha\ra$ (see legend). The Rabi frequency is 
$\Omega= 0.1008$, which suppresses non-resonant transitions as far as 
possible.}
\label{S:fidoft}\end{figure}

In this section, we are mainly concerned with unitary errors due to 
non-resonant transitions, and due to phase-errors which are not suppressed by 
the $2\pi k$-method. Thus, we will study in some detail the behaviour of 
fidelity during the algorithm. The fidelity~\cite{fidrev}, $f(t)$, is the 
overlap between the state evolving under the ideal evolution (where only 
resonant transitions are taken into account) and the state evolving under the 
true evolution governed by the Hamiltonian in (\ref{I:defH}). The fidelity 
(or fidelity amplitude) is a complex quantity. Thus, we will normally study 
its absolute value squared, $|f(t)|^2$.

Figure~\ref{S:fidoft} shows $|f(t)|^2$ during the whole search algorithm for
four different target states, selected by different oracle 
pulses: red curve ($\alpha =0$), blue curve ($\alpha= 5$), light blue curve 
($\alpha= 8$), and pink curve ($\alpha= 13$). In these simulations, we set the 
Rabi frequency to $\Omega= 0.1008$, which corresponds to the $2\pi k$-method 
with $k=4$. Since $k$ is even, non-resonant transitions are suppressed during
$\pi$-pulses, as well as $\pi/2$-pulses.

\begin{figure}
\input{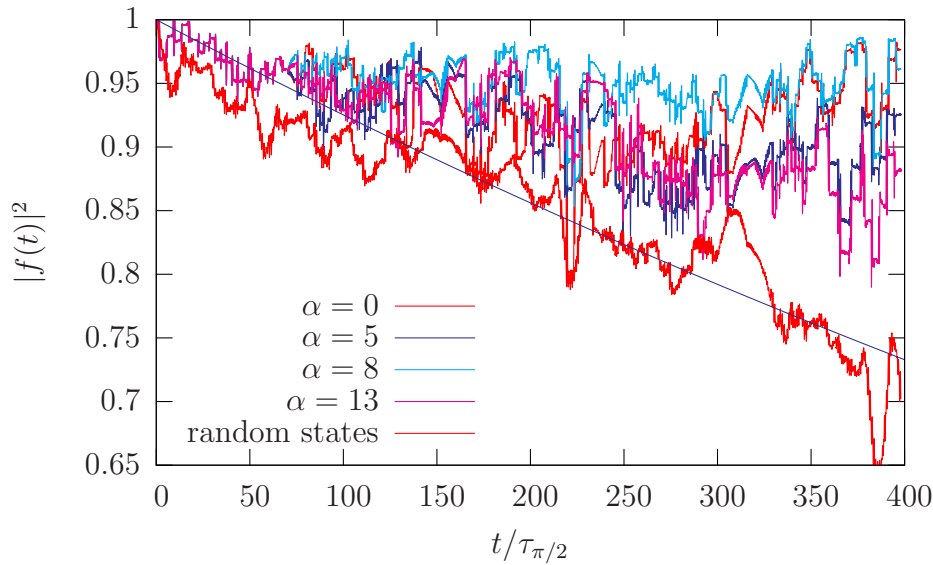}
\caption{The absolute value squared of the fidelity during the Grover 
algorithm with two Grover operator steps. Time is measured in units of the 
duration of a single $\pi/2$-pulse. The algorithm is simulated for different 
target states $|\alpha\ra$ (see legend). The Rabi frequency is 
$\Omega= 0.135$, which yields similarly high fidelities as in 
figure~\ref{S:fidoft}. The smooth blue line shows an exponential, fitted to the
initial decay ($t/\tau_{\pi/2} < 70$). The fastest decay, roughly following 
the exponential is obtained for random initial states (lowest red curve).}
\label{S:fidoft2}\end{figure}

Figure~\ref{S:fidoft2} shows $|f(t)|^2$ for the same Grover algorithm, but for 
the Rabi frequency, $\Omega= 0.135$, which corresponds to the $2\pi k$-method
with $k=3$. Since $k$ is odd in this case, one would expect that non-resonant
transitions are suppressed during $\pi$-pulses, but not during $\pi/2$-pulses. 
Still, we find similarly high fidelities as in the previous case. This 
surprising result is discussed in more detail, below. 
In this figure, the fidelity along the full algorithm does not follow the 
initial trend of relatively fast decay. To obtain that initial trend, we fitted 
an exponential to the initial preparation step (the first 70 $\pi/2$-pulses).

This non-generic behaviour is interesting, because it shows that it is not
sufficient to characterise the performance of individual quantum gates by
their average fidelity-loss rates. With no other information than such loss
rates, one would greatly underestimate the overall fidelity of the algorithm.
A convenient way to obtain average loss rates, is to use random initial states.
The lower red curve shows $|f(t)|^2$ averaged over 100 different initial 
states chosen at random from the invariant (under unitary transformations) 
ensemble of normalised states. The curve still shows large fluctuations which
should disappear in the limit of a larger sample size. However, the general 
trend of a global exponential decay is clearly observable. It also agrees
quite nicely with the exponential fit (smooth blue curve).

\begin{figure}
\input{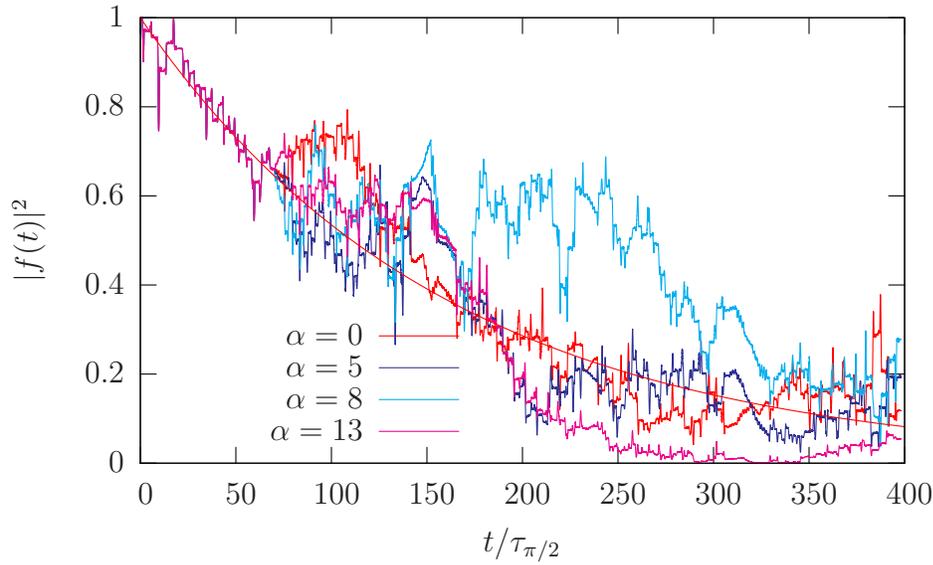}
\caption{The fidelity (its absolute value squared) during the Grover algorithm 
with two Grover operator steps. Time is measured in units of the duration of a 
$\pi/2$-pulse. The algorithm is simulated for different target states 
$|\alpha\ra$ (see legend).}
\label{S:fidoft3}\end{figure}

In figure~\ref{S:fidoft3} the Rabi frequency is chosen to allow non-resonant 
transitions, $\Omega= 0.2550$. By consequence, we find a much faster decay of 
the fidelity. In distinction to the previous figures, the fidelity now follows 
much closer an exponential decay, although there are still very large
fluctuations between the different cases (different oracles).

\begin{figure}
\input{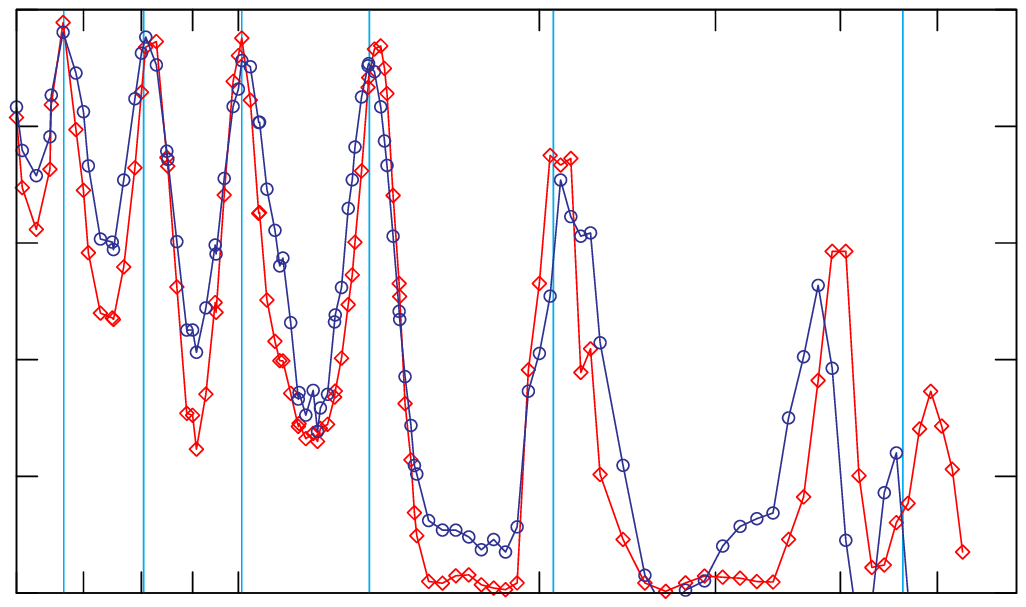}
\caption{The fidelity (blue line) as well as its absolute value 
squared (red line) of the whole Grover algorithm as a function of the Rabi 
frequency, $\Omega$. The algorithm is simulated for the target state 
$|\alpha\ra,\, \alpha= 13$.}
\label{S:fidofrabi}\end{figure}

In figure.~\ref{S:fidofrabi} we plot the real part of the fidelity (blue line) as 
well as its absolute value squared (red line) for the whole algorithm with two 
Grover operator steps as a function of the Rabi frequency, $\Omega$. We chose
the oracle $O(13)$, which selects the target state $|\alpha\ra$ with 
$\alpha= 13$. One can clearly see pronounced peeks of maximal fidelity, where 
the $2\pi k$-method is at work. The Rabi frequencies obtained from the 
condition of the $2\pi k$-method are indicated by vertical lines, where $k$
increases from $k=1$ near $\Omega\approx 0.45$ up to $k= 6$ near
$\Omega\approx 0.067$. 

For integer $k$, the $2\pi k$-method suppresses non-resonant transitions only 
for $\pi$-pulses. In order to suppress non-resonant transitions also for 
$\pi/2$-pulses, $k$ must be even. Since the present algorithm consists mainly
of $\pi/2$-pulses, we should expect rather small fidelities at Rabi 
frequencies which correspond to odd values for $k$. Nevertheless, even for odd
values of $k$, we find pronounced maxima of maximal fidelity. They are almost
as high as the maxima corresponding to even $k$'s. Somehow, it 
happens that non-resonant transitions which occur during one $\pi/2$-pulse are 
undone during a second $\pi/2$-pulse. While this might not be so surprising in 
the case of the $\pi$-pulses which have been split deliberately into 
$\pi/2$-pulses, it is more surprising in the case of $R_y(\pi/2)$ which also 
forms a part of the Hadamard gate [see equation~(\ref{G:Hgen})], and which is 
implemented with ``true'' $\pi/2$-pulses.

\section{\label{C} Conclusion}

We implemented Grover's quantum search algorithm on a Ising spin chain of
length four, taking into account first and second neighbour couplings. The
coupling strength of second neighbours is much weaker than that of first 
neighbours, which means that the dominant source for errors are near resonant
transitions, where the detuning is of the order of $J'$. In other words, the
second neighbour couplings are the dominant source for errors in our setup.
The accumulated error in the course of the quantum algorithm are quantified
by the fidelity. We investigated its temporal behaviour as well as its final
value, at the end of the algorithm, as a function of the Rabi frequency. 

We found that by properly rearranging the RF-pulses, the errors in the 
implementation of the Hadamard gates are greatly reduced. By properly choosing
the Rabi frequency we obtained fidelities beyond $90\%$, for the whole 
algorithm with two Grover steps. This corresponds to an average loss of 
fidelity of the order of $4\times 10^{-4}$ (per $\pi$-pulse) and 
$6\times 10^{-3}$ (per Hadamard gate), numbers which are compatible with recent 
requirements for fault tolerant quantum computation~\cite{knill04prep}. 
However, we also found that the loss of fidelity in different subsections of 
the algorithm varies considerably (sometimes even in sign), which means that 
it is in general not possible to estimate the fidelity of the whole algorithm, 
knowing the fidelities of its parts.

In the present work, we implemented the $2\pi k$-method to suppress non-resonant
transitions. Since the big majority of the RF-pulses used are $\pi/2$-pulses,
we would have expected high fidelities at Rabi frequencies corresponding to
even $k$, but rather low fidelities at Rabi frequencies corresponding to odd
$k$. Surprisingly, we found similarly pronounced maxima of the fidelity at
Rabi frequencies corresponding to even and odd values for $k$.

Preliminary studies using fluctuating Larmor frequencies (simulating tiny
variations of the external magnetic field) indicate that these findings are
not sensitive to external noise. A more detailed study of the effect of noise
will be the subject of future investigations.

%\bibliographystyle{unsrt}
%\bibliography{amol,deco,ranh,semic,stas,books,qcom}

\end{document}